# Random Segmentation: New Traffic Obfuscation against Packet-Size-Based Side-Channel Attacks

Mnassar Alyami [1,2,*] , Abdulmajeed Alghamdi [1], Mohammed A. Alkhowaiter [1], Cliff Zou [1]  and Yan Solihin [1]

[1] College of Engineering and Computer Science, University of Central Florida, Orlando, FL 32816, USA; abdulmajeed.alghamdi@ucf.edu (A.A.); mo894398@ucf.edu (M.A.A.); changchun.zou@ucf.edu (C.Z.); yan.solihin@ucf.edu (Y.S.)

[2] College of Computer Science and Information Technology, Jazan University, Jazan 82822 – 6694, Saudi Arabia

* Correspondence: mnassar.alyami@ucf.edu or malsaad@jazanu.edu.sa

**Abstract:** Despite encryption, the packet size is still visible, enabling observers to infer private information in the Internet of Things (IoT) environment (e.g., IoT device identification). Packet padding obfuscates packet-length characteristics with a high data overhead because it relies on adding noise to the data. This paper proposes a more data-efficient approach that randomizes packet sizes without adding noise. We achieve this by splitting large TCP segments into random-sized chunks; hence, the packet length distribution is obfuscated without adding noise data. Our client–server implementation using TCP sockets demonstrates the feasibility of our approach at the application level. We realize our packet size control by adjusting two local socket-programming parameters. First, we enable the TCP_NODELAY option to send out each packet with our specified length. Second, we downsize the sending buffer to prevent the sender from pushing out more data than can be received, which could disable our control of the packet sizes. We simulate our defense on a network trace of four IoT devices and show a reduction in device classification accuracy from 98% to 63%, close to random guessing. Meanwhile, the real-world data transmission experiments show that the added latency is reasonable, less than 21%, while the added packet header overhead is only about 5%.

## 1. Introduction

The wide adoption of IoT devices comes with a privacy threat. Even with encryption, the metadata of encrypted traffic, such as the packet size, data volume, and packet inter-arrival time, can be utilized by passive observers to conduct *device fingerprinting* (DF) attacks [1,2]. These attacks enable observers to identify the presence of devices and their operational states, thereby allowing adversaries to infer privacy-sensitive information about user behaviors and activities. For example, Wang et al. [3] show that an observer can identify which command a user gives to a smart speaker using the packet length sequence and direction.

DF becomes feasible due to correlated information in encrypted traffic associated with IoT devices. Several studies [4,5] have validated that an observer can passively capture the network traffic and use features of packets' lengths and timing to build machine learning-based classifiers for device identification. Once a device is successfully fingerprinted, the adversary may monitor the fluctuations in the device traffic to detect network events (e.g., Nest Thermostat is in Active or Idle mode) [6]. Hence, protecting against device identification would not only prevent DF, but also hinder event detection (i.e., event-level adversaries must identify the device first and then monitor for status-indicating patterns).

WiFi observers can passively capture network traffic transmitted over the WiFi channel without joining the network (see Section 3). Furthermore, secured WiFi encryption cannot hide the MAC-layer traffic metadata, including the frame size, observation timestamp, and signal strength. The signal strength was not found to be a useful attribute for DF [2].

Thus, fingerprinting defense approaches aim to mutate the lengths and/or transmission time. For example, to address the packet-size leakage, the current traffic shaping methods mainly pad packets with additional bytes to obscure the related characteristics [7]. Regarding the timing side-channel, which falls outside the scope of this paper, adding a random packet delay has been employed as a means to prevent such information leakage [8]. Obviously, both countermeasures introduce data and time overhead.

There has been active research on improving privacy with minimum data overhead [9,10]. These methods are typically centered on minimizing the injected noisy data to conceal IoT traffic. However, these improvements fail to balance privacy protection and overhead [11] (i.e., the attack accuracy or data overhead is high). Inspired by the principles of TCP segmentation [12], we propose an alternative approach to distort length-based patterns without adding noise, thus achieving anonymity with a significantly lower data overhead. Our defensive strategy randomizes the packet lengths by breaking the data stream into random-sized chunks instead of injecting noise data for packet-size obfuscation.

We implement our approach at the application level using TCP socket programming, which makes our defense easier to deploy. In this way, an IoT device manufacturer needs a simple software update on its devices to deploy the proposed defense without changing the devices' operating system or low-level codes. We realize our packet-size control by adjusting two local socket-programming parameters. First, we enable the TCP_NODELAY option to force the operating system to push out each packet with our specified length without waiting for additional data. TCP_NODELAY is a TCP socket option that can be used to turn on/off Nagle's algorithm [13], which by default adds a small latency to improve the network efficiency. It minimizes the number of small TCP segments sent over the network by buffering the data and combining them into larger segments for transmission. Second, we downsize the sending buffer of the socket to prevent the sender from pushing out more data than can be received, which could disable our control of packet sizes.

This paper argues that added noise traffic is needed only to mask data-volume-related features, which is non-discriminatory for devices with highly variable data rates (see Section 5.1.1). In fact, the data volume has been utilized for event detection of an already identified device, the step that our defense prevents in the first place. Thus, noise traffic is often unnecessary to hide device-level signatures, and hence, randomization can be achieved without noise, as proposed in this work.

In summary, we present a new defense against packet-size leakage attacks with the following properties:

- Data-efficient: Traditional countermeasures add noise traffic to hide packet-size-based signatures, resulting in a significant data overhead. Our defense thwarts such leakage without adding any noise; thus, it is much more efficient than noise-based solutions.
- Adaptable: Effective techniques in the literature utilize a fixed dynamic for obfuscation (e.g., padding to the maximum transmission unit (MTU)), which poses a non-optimizable overhead. Our defense utilizes adjustable parameters within the application code, enabling greater flexibility and programmability in managing defense strength and overhead.

The remainder of this paper is organized as follows: We examine the relevant literature and previous studies in Section 2. In Section 3, we present the threat model. Section 4 provides a detailed explanation of our approach for traffic obfuscation. We evaluate our technique and discuss our results in Sections 5 and 6, respectively. Finally, we conclude and discuss future work in Section 7.

## 2. Related Work

Many studies have shown how packet-length information can be exploited to identify IoT devices [4] and specific events [3,6]. The Onion Router (Tor), a well-known privacy-preserving system, addresses such side-channel leakage by sending data in a fixed packet length [14]. Nevertheless, adopting Tor increases the amount of received traffic and adds additional latency due to the multi-hop nature of Tor.

Packet padding has an acceptable effectiveness but incurs a high data overhead. The authors of [7] reported that several packet-padding strategies could thwart the attackers' classification but increased the amount of data sent significantly (>500%). A lightweight solution presented by Pinheiro et al. [15] could reduce the accuracy of IoT device identification to higher than random guessing by 15%. Their mechanism inserts random bytes between 1 and the available space to fill the packet (i.e., to equal MTU). Still, the added noisy data (54%) can lead to an undesirable communication overhead.

A closely related defense [16] can successfully defeat analytics based on WiFi eavesdropping. It uses dummy traffic to shape a pair of devices' traffic to be similar. The technique could spoof other devices' traffic by constructing the flow of dummy packets using prerecorded traces of the targeted device. Thus, an attacker cannot identify a specific device. Moreover, it incurs zero Internet bandwidth overhead by dropping the dummy packets at the access point (AP) before sending them to the Internet. The WiFi-based AP needs to be modified to drop dummy packets, which are flagged using the reserved bit flag on the IP header. However, its effectiveness diminishes when the attacker can monitor the IP-level traffic. Insiders accessing the IP header, such as rogue APs and network snoopers, can filter out flagged dummy packets. As a result, network-layer observers can recover the original/undefended traffic and overcome the defense. Our technique addresses the size-based leakage against both internal and external adversaries (i.e., IP- and MAC-level observers) because our segmentation occurs at the transport layer before the packet construction.

Traffic splitting was initially introduced for multi-path routing [17], which splits the flow across different paths to prevent malicious intermediary nodes from recording the whole traffic. It presented two levels of defense. First, the network-layer defense applies a multipathing strategy within the Tor network to obscure the traffic patterns. The second application-layer defense follows the same concept. It decomposes HTTP requests into subrequests in parallel over multiple paths or sends a single HTTP request for different web objects over different entry nodes in Tor circuits. Assuming partial data is insufficient to perform traffic analysis attacks, this defense is effective against remote observers. However, in this paper, we are also concerned about local eavesdroppers who are physically close to the device transmission range. No middleboxes are involved in this scenario; hence, the attacker can collect the complete capture to perform the attack. On the other hand, our defense considers all observers positioned in the link between the source and destination, remote and local observers alike.

Traffic shaping has been introduced as a routing-optimization technique for vehicular ad hoc networks [18]. This approach employs reinforcement learning to enhance the efficiency of routing decisions, particularly in demanding and real-world situations marked by unstable connections, varying communication ranges, and rapid topology changes. This objective is achieved through distributed reinforcement learning, enabling the routing protocol to learn from vehicle experiences to make optimal decisions and adjust to the network's unpredictable fluctuations. This work cannot be extended to determine packet sizes, as its primary objective is to manage packet routing rather than packet sizes.

Signal-jamming approaches can serve as a defense against traffic analysis attacks in wireless networks, all without the need for adding dummy packets or intentional delays. Generally, this method employs antennas to disrupt traffic at possible adversary positions, effectively elevating the noise level [19]. However, this tactic generates interference that impairs the performance of nearby networks and, furthermore, it is considered unlawful (https://www.fcc.gov/general/jammer-enforcement (accessed on 15 May 2023)).

Packet-size randomization has previously been proposed to address the side-channel leakage in secure shell (SSH) communications [20–22], widely used for secure remote access and communication. However, their proposed modifications are specific to the SSH protocol. Many IoT devices often rely on lightweight messaging protocols to fulfill the IoT communication requirements [23], such as Hypertext Transfer Protocol Secure (HTTPS),

Message Queuing Telemetry Transport (MQTT), Constrained Application Protocol (CoAP), Extensible Messaging and Presence Protocol (XMPP), and Data Distribution Service (DDS). Our work proposes a novel use of random segmentation at the transport layer, responsible for passing the data received from all the application layer communication protocols mentioned above. In this manner, our defense is less demanding for deployment and suitable for the IoT architecture.

TCP segmentation was previously proposed to reduce the per-packet overhead on host processors for wired networks [12]. This approach delays segmenting the data into smaller units and sends it as a larger TCP segment to improve efficiency. Unlike this work, our technique makes the segmentation random and, thus, unpredictable to evade patterns on side-channel information that can be used to identify IoT devices.

## 3. Threat Model

We consider two observation points an adversary can exploit to collect encrypted WiFi traffic. In both scenarios, the attacker is physically located within the signal range of the victim's WiFi router or AP. The attacker can be one of the following:

*Active Observer:* The attacker can set a rogue AP with the same network name as the victim's network, which may lure IoT devices to connect to the rogue AP instead of the legitimate one. In this case, the attacker can observe and analyze the IP-level traffic of the connected IoT devices. We assume the observer can inspect the header of IP packets but does not know the device or break the encryption.

*Passive Eavesdropper:* The attacker can listen to the wireless channel to capture the encrypted WiFi traffic using a WiFi card in monitor mode (https://en.wikipedia.org/wiki/Monitor_mode (accessed on 15 May 2023)). Eavesdroppers are not required to access or join the network. The attacker is, therefore, very hard to detect.

We assume the attacker can access the same IoT devices as the victim's network. The attacker can collect the encrypted traffic to build a profile that can be used to identify IoT devices with similar traffic patterns. Our adversary aims to infer the device (e.g., doorbell, sleep monitor, etc.) and then monitor for network events based on traffic pattern changes, e.g., a surge in doorbell traffic indicates the arrival of a visitor, a surge in sleep monitor traffic indicates a user is awake, etc. We consider device-fingerprinting attacks that operate on packet lengths and directions. Timing information falls beyond this paper's scope and, therefore, is not considered by our defense.

## 4. Materials and Methods

### 4.1. Noise-Free Randomization

We defend against traffic-analysis attacks by enabling IoT applications to control their packets, where it is not typically controllable. The application layer delivers a message of byte stream to the transport layer, which appends its header information (e.g., port number) and passes the data to the network layer as a segment. The segment size is determined by the maximum segment size (MSS) and specific situations summarized below [24]:

- If the application message $\geq$ MSS, the TCP protocol sends the data in full segments equal to MSS for transmission and holds any portion of data surpassing MSS in an incomplete segment accumulating more bytes.
- If the message $<$ MSS (i.e., there is still space in the segment) and a previously sent packet has not been acknowledged yet, the protocol waits for some time ($+/-$ 200 ms [13]) to accumulate more bytes. This time delay allows for collecting more data to optimize network usage.
- If no additional data arrives within the timer period, the protocol dispatches the available data for transmission.

The network layer, which is responsible for sending the data over the network, receives the segments and breaks them into as few full packets as possible by the MTU restriction. Indeed, the MSS value at the transport layer depends on the underlying network MTU to ensure TCP segments can be properly encapsulated within network packets. MSS parameters are exchanged during the TCP-handshake phase, and the network interface card's device driver provides the TCP/IP stack with the MTU value.

With our defense, the packet size is indistinguishable due to the randomness. That is, it consistently breaks thet received messages into arbitrarily sized segments. Suppose an IoT application sends a message $m$ of $n$ bytes. As stated earlier, the system, by default, will transmit $m$ in a single segment if $n$ = MSS or pass the small message ($n$ < MSS) as it is after the timeout period ($+/- 200$ ms). If $n$ > MSS, $m$ will be divided into $\lceil n/\text{MSS}\rceil$ segments, all equal to MSS except the last segment if $n$ is not an exact multiple of MSS (i.e., $n\%\text{MSS}> 0$); then, the last segment will hold the remaining bytes ($n\%\text{MSS}$). For example, assuming MSS is 1500, an application message of 3500 bytes will be transmitted in three segments (1500, 1500, 500). Unlike this dynamic, our defense will divide $m$ into a random number of segments, and each segment's length will also be random. As a result, the data pattern from the length features is not predictable.

Figure 1 depicts two communication scenarios between a cloud server and an IoT device; the top represents the regular/undefended traffic, and the bottom shows the shaped/defended one. In both examples, we assume the server sends the same message two times. We here focus on the incoming traffic from the server side to demonstrate the concept of our defense. However, we expect both endpoints (i.e., server and client IoT) to implement our defense when communicating with each other. Thus, the bidirectional traffic is fully obfuscated.

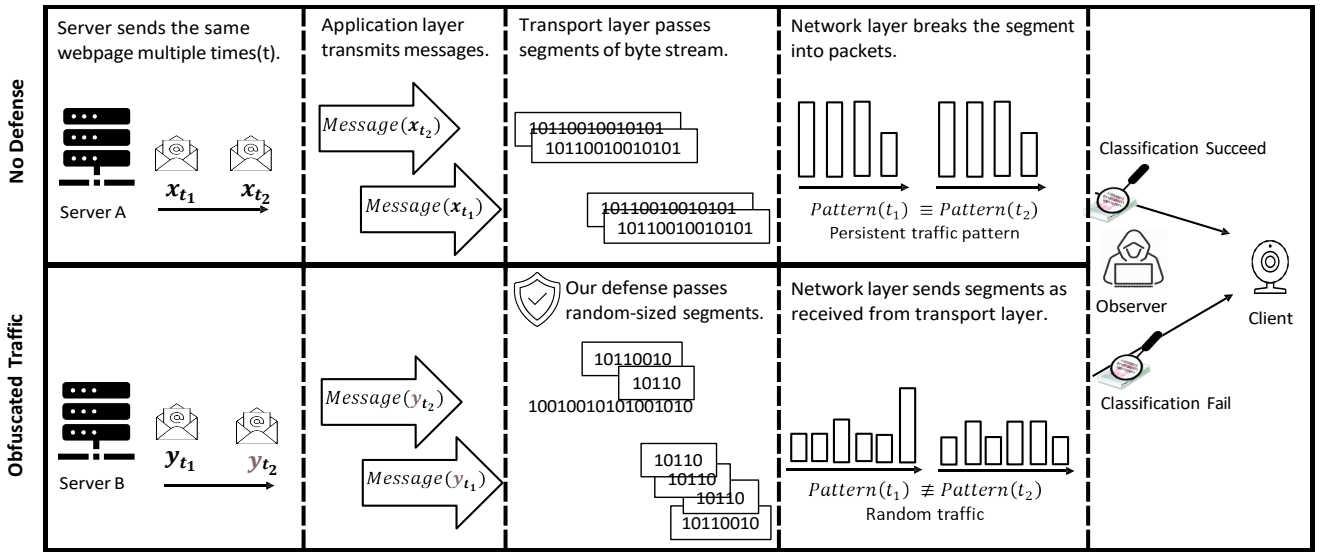

**Figure 1.** Two communication scenarios between a cloud server and a client IoT device. The top shows the original traffic without any defense, and the bottom depicts the obfuscated traffic after utilizing our proposed defense.

In the first scenario, the server with no defense sends data in a typical packetized flow, leaking exploitable signatures for fingerprinting attacks. In contrast, the second defended scenario shows that randomization occurs at the transport layer. Specifically, our technique passes random-sized segments to the network layer. Consequently, the traffic pattern generated by our shaping technique is challenging to classify.

### 4.2. Algorithm

Algorithm 1 shows the pseudo-code processes of the proposed defense. We assume cloud servers and IoT devices run our program when sending packets.

---

**Algorithm 1** Random Segmentation of Application Messages

---

1: *Data*[] ← Byte array holding the application message.
2: *Min*, *Max* ← Select the minimum and maximum segment sizes.
3: *Prob* ← Select the segmentation probability.
4: **if** length of *Data* ≥ *Min* **and** *random.random*() ≤ *Prob* **then**
5:    *start* ← 0
6:    *end* ← length of *Data* − 1
7:    **while** *start* ≤ *end* **do**
8:      *RandLen* ← random(Min, Max)
9:      **if** *start* + *RandLen* ≥ *end* **then**
10:        *index* ← *end*
11:      **else**
12:        *index* ← *start* + *RandLen*
13:      **end if**
14:      *Seg* ← *Data*[*start* : *index*]
15:      **send** *Seg*
16:      *start* ← *index* + +
17:    **end while**
18: **else**
19:    *Data* are left to be handled by the operating system without segmentation.
20: **end if**

---

The program stores the received message from the application layer in a byte array *Data*. Provided that the message is long enough to perform random segmentation (such that the length of *Data* ≥ the minimum segment size *Min*), our algorithm randomly decides whether to split the message but with a specific probability threshold *Prob*, such that 0 ≤ *Prob* ≤ 1. If yes, the main loop loads a random chunk of *Data* into an individual segment *Seg* for transmission and loads another random chunk from *Data* in the next iteration process until the array becomes empty. The size of each segment is determined randomly by *RandLen*, but does not surpass MTU to avoid fragmentation. However, the upper and lower bound of *RandLen* (i.e., *Min* and *Max*) is adjustable to suit the device traffic pattern. For example, for a device that sends light traffic with a maximum of 300 bytes in length, the upper bound of *RandLen* should be less than 300 to achieve randomness. Otherwise, the program will send the whole payload in *Data* if *RandLen* exceeds the message's size.

### 4.3. Multi-Level Segmentation

Due to the significant differences in the packet size range of many IoT devices operating in different modes, choosing the appropriate range for length randomization (i.e., *Min* and *Max*) is challenging. For example, our preliminary analysis of our camera traffic shows that 93% of the packets are below 150 bytes when the camera is idle. On the other hand, when the camera becomes active, 58% of packets are above 1000 bytes. Hence, utilizing one range to mask all functional scenarios, such as splitting all segments into chunks between 100 and 150 bytes, will obfuscate the whole traffic but lead to excessive segmentation and increase the overhead.

Given the limitation of the one-level segmentation, it is necessary to make our defense adapt to the change in traffic volume. Thus, we adopt a multi-level segmentation to enable our algorithm to use a suitable range based on the traffic intensity. For example, we use three levels for the high-bandwidth devices. Level 1 splits messages ≤200 bytes into random chunks between 20 and 40. Other messages above 200 and ≤500 are randomized using random lengths between 100 and 300 in level 2. Larger streams in level 3 that are above 500 are obfuscated using a random size between 500 and 1000. Obviously, that creates non-overlapping bands where the observer can recognize the corresponding band, but it is still insufficient to perform the attack due to the randomness within each band.

#### 4.4. Practical Considerations

To change the standard packet sizing enforced by underlying protocols controlled by the operating system, we initially considered changing the MTU. Dynamic modification of the MTU value will directly impact the packet size, breaking larger packets into smaller ones to fit within the new limit. However, MTU is a system-wide parameter, and such modification will not affect our program but the entire system, which is undesirable. In addition, smaller MTUs increase packet fragmentation, leading to adverse consequences for the system's performance. Assembling fragments at the destination adds extra burden and reduces the overall efficiency. Further delay can also occur when a fragment is missed or corrupted. In this case, the receiver cannot read partial data, and the whole data frame must be retransmitted. On the other hand, our approach does not fragment packets but divides TCP segments into separate IP packets. Hence, we avoid the drawbacks associated with packet fragmentation.

The main challenge in applying our idea is that applications do not have direct packet abstraction to control packet lengths. We could work around this technical obstacle by modifying two parameters on a per-socket basis, thus not affecting other programs.

First, we disable Nagle's algorithm [13] using the TCP_NODELAY option. Nagle's algorithm avoids sending small TCP segments by introducing a time delay to collect more data so that it sends full segments. Since we aim to send data in predetermined sizes, this data aggregation contradicts our purpose and needs to be disabled.

Second, in the case of intensive traffic, we limit the amount of data that can be pushed out of a socket at the initial stage of the communication. TCP begins with a small congestion window to assess the network condition and find the optimal window size. As we turn off Nagle's algorithm, many small packets can be sent immediately. As a result, the operating system overrides our program and accumulates the data into larger packets to improve efficiency, rendering our defense ineffective. To avoid this problem, we decrease the send socket buffer to less than the receive buffer.

As stated, our traffic-shaping technique masks the packet length without adding noisy data. Thus, the data rate remains unchanged (i.e., the amount of transmitted data is the same). Therefore, we added another module to select the amount of covered traffic as needed. In our approach, we inject a certain amount of traffic to make one device similar to another in terms of the data rate.

#### 4.5. Implementation

We developed a server–client implementation using socket programming in Python. We use the TCP protocol to send packets due to its reliability and expect our obfuscation methodology to apply to UDP as well. Our source code is publicly available at GitHub (https://github.com/MnassarAlyami/Random-TCP-Segmentation.git (accessed on 29 July 2023)).

We assume the defender has access to the targeted IoT devices and servers to install our program as a patch used by a hook that intercepts every send command from the application layer. That is, it will replace the standard send code with our patch to randomize the packet size.

### 5. Evaluation and Results

In this section, we first evaluate our defense against traffic classification of IoT devices based on packet length. We compare the performance of our noise-free randomization with an analogous noise-based mechanism that uses random packet padding [15]. Second, we quantify the impact of our technique on communication performance through real-world experiments. Below, we discuss each aspect and present our results.

#### 5.1. Effectiveness

To evaluate the randomness in the packet length introduced in our technique, we developed a program to simulate our defense on a WiFi trace of four IoT devices: doorbell, camera, light bulb, and smart plug. We captured the encrypted traffic for one hour in different operating modes (e.g., ON-OFF).

Our program reads the pcap file of each device and produces the obfuscated traffic to test our defense against DF attacks. Table 1 outlines our configuration for the adjustable system parameters introduced in Sections 4.2 and 4.3. The probability was manually chosen with the goal of minimizing the overhead (i.e., reducing the number of segmented packets) while maintaining a lower accuracy. The initial value of 0.6 resulted in a high accuracy (>80). Consequently, we increased the probability to 0.7, and the accuracy remained consistently high. Subsequently, when we further elevated the probability to 0.8, a reduction in accuracy was achieved. We used the same defense parameters for each group of devices categorized based on the traffic intensity because different parameters will likely create new patterns to distinguish the devices.

**Table 1.** Our experimental setup of the system parameters *.

| Parameter | Low-Bandwidth Devices | High-Bandwidth Devices | | |
| --- | --- | --- | --- | --- |
| | **Bulb and Plug** | **Doorbell and Camera** | | |
| *Prob* | 0.8 | 0.8 | | |
| *Min* | 5 | L1 = 20 | L2 = 100 | L3 = 500 |
| *Max* | 20 | L1 = 40 | L2 = 300 | L3 = 1000 |

* L1, L2, and L3 refer to the three levels of segmentation introduced in Section 4.3.

Furthermore, we run another simulation to obfuscate our captured trace using the random padding introduced in [15] to compare the effectiveness of a traditional noise-based solution with our defense.

The attacker's profiling classifier is trained using the training data from the original trace. To assess the attack's performance without defense, we test the classifier using the testing data derived from the original trace.

Likewise, we evaluate the defense approach by initially creating a modified trace using our defense program, relying on the original trace as a foundation. Subsequently, we proceed to train the attack classifier using the training data from the modified trace and then test the classifier using the corresponding test data in the modified trace.

We use Random Forest for our classification due to its outperformance on similar IoT device-identification attacks compared with several other ML algorithms [2,4]. We randomly divided our dataset into 70% for training and 30% for testing and quantified the performance of our classifier using the following metrics: accuracy, precision, recall, and F1 score. The specific computation of these metrics can be found in the Appendix A.

Note that we show the indistinguishability of devices by applying our technique to training and testing data; the closer the accuracy to random guessing (50%), the more effective our defense is in confusing the classifier. Random guessing attains an accuracy rate of $1/k$, where $k$ is the number of labels/devices. As we confuse the attacker between two devices of similar traffic intensity, then $k = 2$.

We evaluate against an attacker who exploits the packet size and direction only. Thus, we construct our dataset using packet sizes in a binary format to represent directions; a positive size represents incoming packets, and a negative size, outgoing ones. Similar to [2,16], we break the trace into sequences observed within a 30 s time window for classification.

### 5.1.1. Preliminary Data Analysis

Figure 2 presents a sample traffic flow observed over a period of 10 min before and after implementing our defense. Before obfuscating the traffic, we observed variable traffic patterns in the light bulb and smart plug. Specifically, the bulb's traffic was higher than the plug's in seven instances, lower in two instances, and similar in one case (Figure 2a). This inconsistency in traffic poses a challenge for reliable device profiling based on data-volume-related features. The impact of this variability is evident in the spike in bulb traffic (Figure 2a, traffic period 5), which was misclassified as camera traffic.

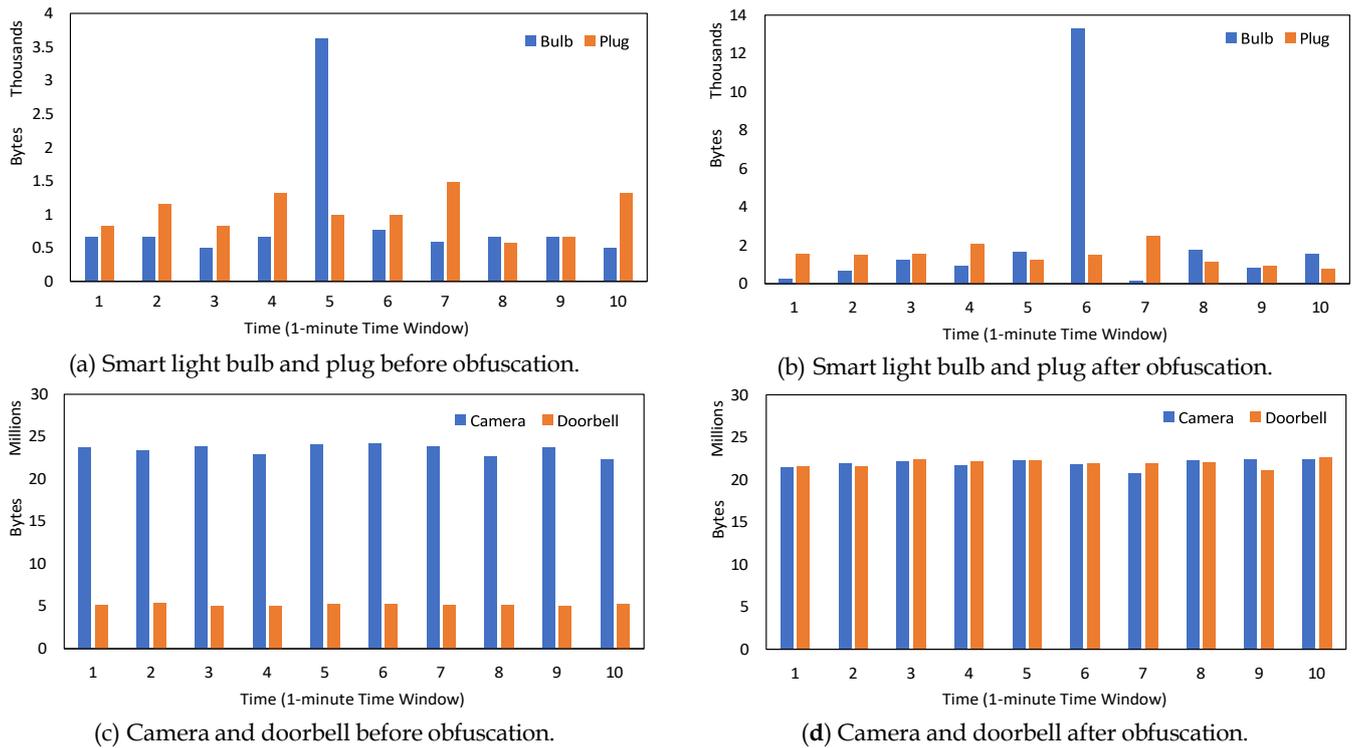

(a) Smart light bulb and plug before obfuscation.

(b) Smart light bulb and plug after obfuscation.

(c) Camera and doorbell before obfuscation.

(d) Camera and doorbell after obfuscation.

**Figure 2.** Traffic flow of four IoT devices over 10 min.

On the contrary, the camera and doorbell exhibit stable traffic patterns while operating in a fixed working mode, such as recording videos (Figure 2c). However, the camera's higher resolution results in significantly larger traffic compared to the doorbell. Therefore, to account for the variance in data volume between the camera and doorbell, we introduced covered bytes to the doorbell to compensate for the variance in data volume (Figure 2d). Dummy/covered packets can be labeled and discarded at the receiving side. We assume the defender can leverage the header field of the Traffic Flow Confidentiality (TFC) mechanism [25]. This mechanism provides a tool to inject dummy packets using a wrapped header field that is encrypted and cannot be observed by network observers. The injection statistics are summarized in Table 2, revealing that no dummy packets were injected for the bulb and plug since the data rate does not serve as a suitable representative for those two devices. Conversely, the doorbell necessitates a greater incorporation of covered bytes to achieve parity with the camera in order to introduce confusion within the classifier's discrimination between the two devices. Additionally, a marginal proportion of covered bytes (0.7%) was introduced to the camera to mask its disparities from the doorbell, particularly during periods of inactivity.

**Table 2.** Injected covered bytes to hide data-rate features.

| Device | Covered Bytes (%) |
|---|---|
| Bulb | 0 |
| Plug | 0 |
| Camera | 0.7 |
| Doorbell | 340 |

It is important to highlight that we incorporated a 20% time delay (refer to Equation (2) for the specific computation) in our simulation to align it with the findings in Section 5.3. As a result, the overall traffic attributes were affected. For instance, the event surge observed in the light bulb's traffic before applying our defense (Figure 2a, traffic period 5) can be seen in the subsequent observation time window in the defended traffic (Figure 2b, traffic period 6).

Figure 3 illustrates the impact of our approach on the packet size. We chose the bulb and plug for demonstration as they rely entirely on random segmentation for packet-size obfuscation (i.e., no covered bytes were injected). Before implementing our defense, we can notice in Figure 3a a steady average size (nearly 130 bytes) sent by the plug versus fluctuating value by the light bulb. (After subtracting the frame header (82 bytes), a data frame of 130 bytes means there are 48 (i.e., 130-82) bytes in the payload for the segmentation.) We can observe similar behavior in the return traffic represented by negative values in Figure 3c. After obfuscating the traffic (Figure 3b,d), the described range of each device starts to overlap, introducing uncertainty in the learning process for device classification.

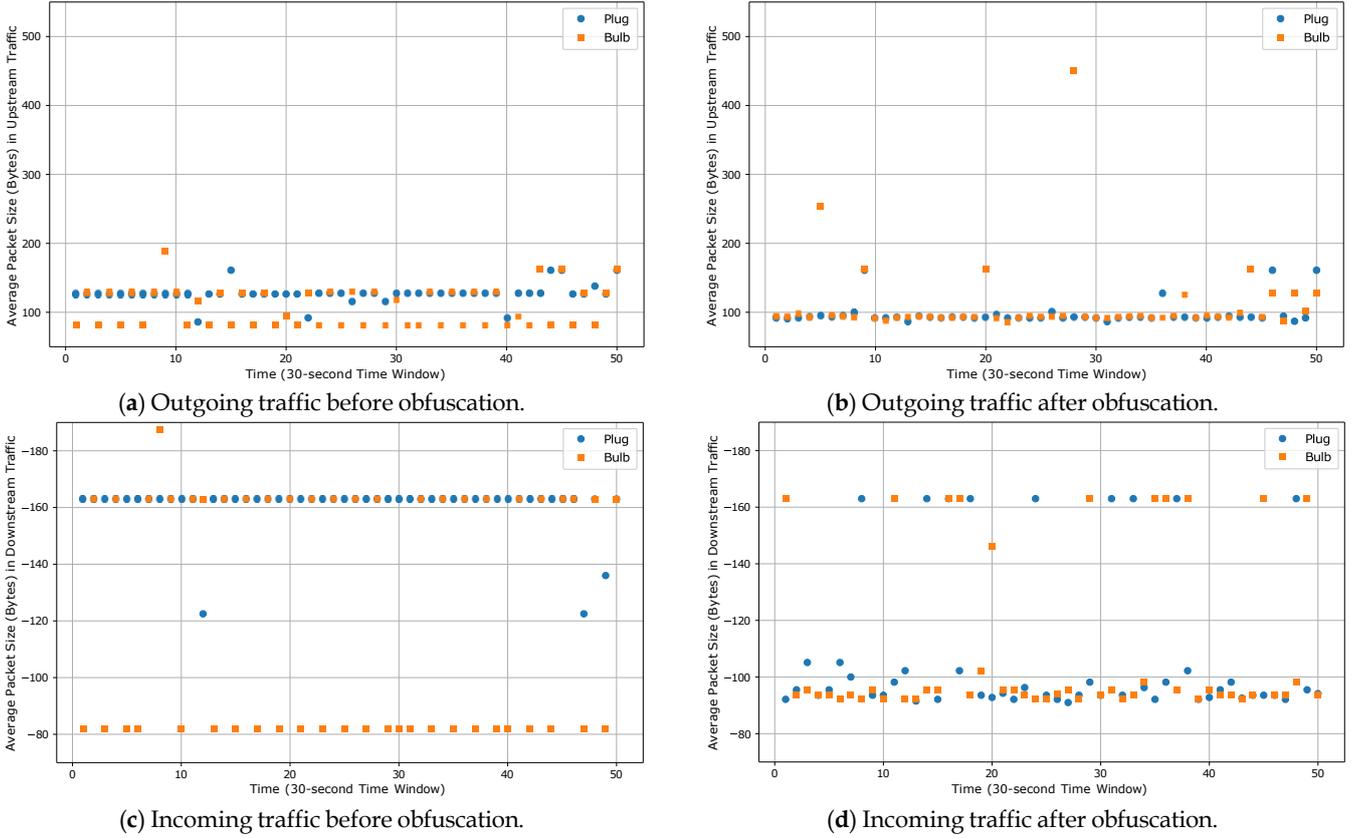

**(a)** Outgoing traffic before obfuscation.

**(b)** Outgoing traffic after obfuscation.

**(c)** Incoming traffic before obfuscation.

**(d)** Incoming traffic after obfuscation.

**Figure 3.** Average packet size before and after obfuscating the bidirectional traffic of two devices over time.

### 5.2. Efficiency

In this section, we evaluate the byte overhead $B$ and the time taken $T$ by our algorithm. We calculate $B$ as:

$$B = \frac{D_b - W_b}{W_b} \quad (1)$$

where $D_b$ is the total amount of bytes transferred when implementing our defense, and $W_b$ is the total amount of bytes transferred without implementing our defense. As we utilize the covered bytes presented in Table 2 solely for concealing data rate features, we deduct them from this calculation in the context of packet-size obfuscation.

For the second aspect ($T$), we set up a remote virtual server and let our local machine send a large file of 10 MB, with and without our defense. Thus, we calculate the added latency by our methodology compared with the standard/undefended transmission scenario. We define $T$ as:

$$T = \frac{D_t - W_t}{W_t} \quad (2)$$

where $D_t$ is the time span to send the file when implementing our defense, and $W_t$ is the time span to send the same file without implementing our defense.

Furthermore, we implemented two randomization levels to analyze the impact of the obfuscation intensity (i.e., range of random values) on $T$. The wider the range of random lengths, the more packets are required to carry the payload. Consequently, more packets might take a longer time to deliver. For instance, sending a large array of data using random-sized packets ranging from 100 to MTU will result in significantly more packets than using a range of larger lengths between 1200 and MTU. For this experiment, we define two randomization levels ($Rand_{(low)}$ and $Rand_{(high)}$) with an upper bound of a common maximum length of 1400 bytes, whereas the lower bound of each level varies as follows: $Rand_{(low)}$ = 1200 and $Rand_{(high)}$ = 100. We run ten sets of experiments and report the average result in the following section.

### 5.3. Results

As shown in Table 3, all the classification metrics used to evaluate the randomness in the shaped traffic are close to the baseline of random guessing. The values in the last column are bolded to represent the best-performing results. The result demonstrates the reduction in classification accuracy from 98% by the attack to 63% by our defense. Also, our technique achieves a better obfuscation (lower accuracy) than random padding. The attack accuracy under our defense is 8% lower.

Moreover, Table 4 compares the byte overhead $B$ of our defense with random padding. The bolded values in the last column represent the most favorable outcomes. Our defense incurs a significantly lower overhead for all devices, which saves nearly 47% of the total overhead. (From Table 4, $B$ with random padding is 54% more than with random segmentation, which achieves 7%, resulting in a savings of 47% (54-7).)

**Table 3.** Classification accuracy, precision, recall and F1 score of two defenses.

| Metric | No Obfuscation (%) | Random Padding [15] (%) | Random Segmentation (%) |
|---|---|---|---|
| Accuracy | 98 | 71 | **63** |
| Precision | 98 | 77 | **67** |
| Recall | 98 | 71 | **63** |
| F1 | 98 | 71 | **63** |

**Table 4.** Byte overhead $B$ of two defenses *.

| Device | No Obfuscation | Random Padding [15] | | Random Segmentation | |
|---|---|---|---|---|---|
| | $W_b$ (MB) | $D_b$ (MB) | $B$ (%) | $D_b$ (MB) | $B$ (%) |
| Bulb | 0.0348 | 0.1936 | 456 | 0.1091 | **214** |
| Plug | 0.0287 | 0.1626 | 467 | 0.0887 | **209** |
| Camera | 549.2 | 766.7 | 40 | 589 | **7** |
| Doorbell | 155.3 | 317.6 | 105 | 168 | **8** |
| Total | 704.6 | 1084.7 | 54 | 757.2 | **7** |

* Refer to Equation (1) for details regarding the definition of $W_b$, $D_b$, and the specific computation of $B$.

Last, we report the latency results from our large file transmission experiments between our client machine and a remote server. As shown in Figure 4, our defense comes with an average time overhead of 20.5%. Compared with random padding, our technique underperforms by only 0.7%, as [15] reported a 19.8% latency. The same figure (Figure 4) also shows a stable $T$, regardless of whether we perform low or high splitting (using $Rand_{(low)}$ and $Rand_{(high)}$). Although $B$ has increased by 4.5% due to the intensive splitting using $Rand_{(high)}$, more splitting seems not to introduce noticeable latency.

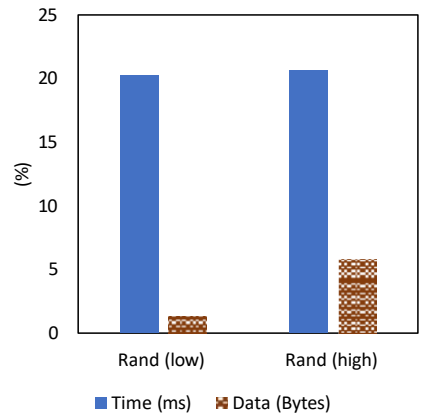

**Figure 4.** Latency (*T*) and packet header overhead (*B*) using the two randomization levels: $Rand_{(low)}$ and $Rand_{(high)}$.

## 6. Discussion

Effectiveness: Our results validate that our approach can disrupt packet length features to protect against traffic-analysis attacks. By simulating our defense on a trace from real IoT devices, we show that the obfuscated traffic resulted in a baseless classifier comparable to random guessing. This is because our defense sends data packets in random sizes, which prevents the classifier from learning length-based fingerprints for profiling.

Our technique does not consider timing characteristics, such as the interarrival time. However, the issue of timing leakage has been addressed by delaying the data packets to obscure the related patterns [8]. This approach can be integrated with our defense to effectively conceal both the timing and the length features.

Efficiency: Our randomization technique is noise-free, and hence, more data-efficient than other noise-based approaches like packet padding. However, there is a case where the header overhead of our defense is higher than padding. If the size of the transmitted flow is relatively small compared to the packet headers (54 bytes), then every split with our approach adds more data than the payload itself. For example, if we have a device that sends a small packet of 100 bytes per second, it becomes expensive to split the packet into multiple chunks, as the header overhead from generating an additional packet becomes 54% (54/100). Nevertheless, such small traffic would be marginal to the total bandwidth in the network.

In terms of time overhead, our countermeasure incurs a reasonable latency (20.5%), which many IoT devices can tolerate, such as sleep monitors and smart plugs [26]. However, high-bandwidth devices like cameras may experience degradation due to the need for greater bandwidth to support video streaming.

One interesting insight we observe in Figure 4 is that intensive obfuscation (i.e., a higher degree of randomness) does not increase the latency. Although the number of packets is higher with our mechanism, it does not add a noticeable latency due to the immediate transmission enabled by the TCP_NODELAY option. The factor that led to the increase in transmission time was the limited send buffer, which puts the socket on hold from pushing more data until the buffer is empty. It is evident from Table 5 that smaller buffer sizes increase the time overhead significantly.

Note that we are not claiming that the deactivation of Nagle's algorithm is an efficient solution, as sending small packets can result in additional header and processing overhead. However, turning off Nagle's algorithm has been introduced in prior studies as an effective technique with no adverse effect on performance, such as preventing many deadlock situations [24]. Similarly, our proof-of-concept implementation on consumer-grade laptops shows that our defense can mitigate privacy leakage but may introduce some latency, as the sender needs to send the data stream in a larger number of packets. With that being said, further research is needed to investigate the impact of our technique on devices with limited processing capacity and storage, similar to IoT devices. We plan further investigations in this direction as future work.

Compatibility and Deployment Challenges:

While our current implementation showcases the feasibility of the approach through a client–server setup using TCP sockets, we fully acknowledge that IoT environments present unique challenges. For instance, adapting our proposed technique for communication with generic browsers in server roles may necessitate updates on server-side software. Hence, there could be technical issues in accommodating all existing devices, necessitating further investigation in future research.

Vulnerability Analysis: It is vital to address potential vulnerabilities and ensure the robustness of our approach. However, our technique only enables IoT applications to change the packet length without any other modification of the entire IoT device's communication. For example, it does not affect WiFi encryption, IPsec, or SSL implementation, etc. Hence, we see no immediate and evident vulnerabilities in our proposed method.

Adversarial Attack: We assume the attacker knows how our defense works and, hence, can try to merge the length of consecutive packets to overcome our defense. However, there is no basis for the attacker to retrieve packet-size patterns. If the attacker combines all consecutive packets, the attacker will merge packets that were not initially split because it is customary to observe a series of small-sized packets, such as mouse flow, when there is no defense implemented. In addition, our mechanism performs random segmentation for randomly selected messages. Hence, there is no fixed rule for our splitting to perform adversarial de-splitting with high accuracy.

**Table 5.** Send buffer size impact on transmission time using two obfuscation levels.

| Buffer Size | Time Overhead (%) | |
| --- | --- | --- |
| | $Rand_{(Low)}$ | $Rand_{(High)}$ |
| $2^{15}$ Bytes | 132.53 | 128.67 |
| $2^{16}$ Bytes | 26.1 | 26.78 |

## 7. Conclusions and Future Work

In this paper, we have shown how random segmentation can obfuscate packet-size patterns without introducing additional noise into the packets themselves, as is the case with packet padding. The proposed approach enables network devices to send application messages through random-sized segments and pass them to the network layer for immediate transmission. Therefore, the observed traffic at and above the network layer is randomized, defending against both in-network and out-network observers (i.e., IP- and MAC-level observation). The technique has been tested on a client machine connected to a remote server, and the results demonstrate the effectiveness of our defense with a reasonable time overhead (<21%).

For future work, we seek to make our defense accommodate the heterogeneity in the IoT environment. The adjustable parameters in our code allow the defender to choose the suitable obfuscation level to achieve sufficient randomness with fewer splitting operations. Thus, our defense system lacks an adaptive functionality to adjust its parameters based on the device traffic intensity and specific hardware. To this end, we aim to present an optimization model to enable our system to dynamically choose the optimum parameters that yield astonishingly less overhead with maximum privacy protection.

**Author Contributions:** Conceptualization, M.A., C.Z., and Y.S.; methodology, M.A. and C.Z.; software, M.A.; validation, M.A., C.Z., and Y.S.; formal analysis, M.A., C.Z., and Y.S.; investigation, M.A., C.Z., and Y.S.; resources, M.A., C.Z., and Y.S.; data curation, M.A.; writing—original draft preparation, M.A.; writing—review and editing, M.A., A.A., M.A.A., C.Z., and Y.S.; visualization, M.A.; supervision, C.Z. and Y.S.; project administration, C.Z. and Y.S.; funding acquisition, C.Z. and Y.S. All authors have read and agreed to the published version of the manuscript.

**Funding:** This research was sponsored by the U.S. National Science Foundation (NSF) under Grant DGE-2325452.

**Data Availability Statement:** Data are available upon request from the corresponding author.

**Conflicts of Interest:** The authors declare no conflicts of interest.

## Abbreviations

The following abbreviations are used in this manuscript:

| | |
|---|---|
| IoT | Internet of Things |
| DF | Device Fingerprinting |
| Tor | The Onion Router |
| AP | Access Point |
| MTU | Maximum Transmission Unit |
| MSS | Maximum Segment Size |

## Appendix A

We employed four metrics to evaluate the effectiveness of our defense: accuracy, precision, recall, and F1 score. These metrics can be computed using the following formulas:

$$Accuracy = \frac{T}{T + F} \tag{A1}$$

where $T$ is the number of instances that are correctly classified and $F$ denotes the instances that are incorrectly classified by the model.

$$Precision = \frac{TP}{TP + FP} \tag{A2}$$

$$Recall = \frac{TP}{TP + FN} \tag{A3}$$

where $TP$ is true positives, $TN$ is true negatives, $FP$ is false positives, and $FN$ is false negatives.

The F1 score calculates the harmonic mean between the precision and recall as:

$$F1 = 2 \times \frac{Precision \times Recall}{Precision + Recall} \tag{A4}$$